\documentclass[%
apl,
reprint,
superscriptaddress,
amsmath,amssymb,
]{revtex4-1}

\usepackage{graphicx}
\usepackage{dcolumn}
\usepackage{bm}
\usepackage{color}
\usepackage[pdfborderstyle={/S/U/W 1}]{hyperref}
\begin{document}

\preprint{}

\title{High spatial coherence in multiphoton-photoemitted electron beams}

\author{Stefan Meier}
\email{stefan.m.meier@fau.de}
\affiliation{Department of Physics, Friedrich-Alexander-Universit\"at Erlangen-N\"urnberg, Staudtstrasse 1, D-91058 Erlangen, Germany, EU}
\author{Takuya Higuchi}%
\affiliation{Department of Physics, Friedrich-Alexander-Universit\"at Erlangen-N\"urnberg, Staudtstrasse 1, D-91058 Erlangen, Germany, EU}
\author{Manuel Nutz}%
\affiliation{Fakult\"at f\"ur Physik and Center for NanoScience (CeNS), Ludwig-Maximilians-Universit\"at M\"unchen, Geschwister-Scholl-Platz 1, D-80539 M\"unchen, Germany, EU}
\author{Alexander H\"ogele}%
\affiliation{Fakult\"at f\"ur Physik and Center for NanoScience (CeNS), Ludwig-Maximilians-Universit\"at M\"unchen, Geschwister-Scholl-Platz 1, D-80539 M\"unchen, Germany, EU}
\author{Peter Hommelhoff}%
\email{peter.hommelhoff@fau.de}
\affiliation{Department of Physics, Friedrich-Alexander-Universit\"at Erlangen-N\"urnberg, Staudtstrasse 1, D-91058 Erlangen, Germany, EU}

\date{\today}

\begin{abstract}
Nanometer-sharp metallic tips are known to be excellent electron emitters. They are used in highest-resolution electron microscopes in cold field emission mode to generate the most coherent electron beam in continuous-wave operation. For time-resolved operation, sharp metal needle tips have recently been triggered with femtosecond laser pulses. We show here that electrons emitted with near-infrared femtosecond laser pulses at laser oscillator repetition rates show the same spatial coherence properties as electrons in cold field emission mode in cw operation. From electron interference fringes, obtained with the help of a carbon nanotube biprism beam splitter, we deduce a virtual source size of less than $(0.65\pm0.06)\,$nm for both operation modes, a factor of ten smaller than the geometrical source size. These results bear promise for ultrafast electron diffraction, ultrafast electron microscopy and other techniques relying on highly coherent and ultrafast electron beams.\\
\textit{The corresponding Applied Physics Letters paper is available at:\\ \href{https://doi.org/10.1063/1.5045282}{https://doi.org/10.1063/1.5045282}}.\\
\end{abstract}


\maketitle
The ability of electrons to interfere, given by their coherence properties, enables matter wave experiments with electrons, such as diffraction, interference or electron holography, as well as highest resolution microscopy \cite{hasselbach2010}. These techniques all rely on highly coherent electron sources. In recent years, great efforts have been undertaken to equip these techniques also with high {\it temporal} resolution. Applications like ultrafast electron diffraction \cite{Zewail2006,Aidelsburger2010} or ultrafast electron microscopy \cite{Barwick2009,Feist2017} are only a few examples. Spatially coherent and ultrafast pulsed electron sources are required for these applications. Laser-triggered electron sources such as flat photocathodes have been employed. More recently, electron emission with high spatial coherence {\it and} high temporal resolution down to femtosecond timescales has been reached by triggering the emission from a metallic nanotip with ultrashort laser pulses \cite{Hommelhoff2006,Hommelhoff2006_2, Ropers2007,Yanagisawa2009,Bormann2010,Schenk2010,Krger2011,Kealhofer2012,Herink2012,Gulde2014,Juffmann2015,Vogelsang2015,Krger2018}. Even sub-femtosecond control has been shown using the carrier-envelope phase of the exciting few-cycle laser pulse \cite{Krger2011,Piglosiewicz2013}. Besides the fundamental investigations on this topic, ultrafast electron beams from needle tips have been already used in initial experiments \cite{Mller2014,Bainbridge2016}. In both cases, a sample is optically pumped with a femtosecond optical pulse and then probed with an ultrashort electron pulse, generated from a second femtosecond optical pulse focused on a metallic nanotip. However both experiments rather rely on a particle projection image of the electron beam and do not seem to take advantage of the electrons' coherence yet. For completeness, we mention a low-energy electron diffraction experiment that does take advantage of electron coherence, but there longer timescales, quite complex electron optics and likely a filtered electron beam have been employed \cite{Gulde2014}. For forthcoming experiments on ultrafast electron diffraction or holography, it is highly advantageous to attain and to investigate the spatial coherence of ultrafast electron beam sources as done here.

\begin{figure}[t!]
	\begin{center}
		\includegraphics[width=8.5cm]{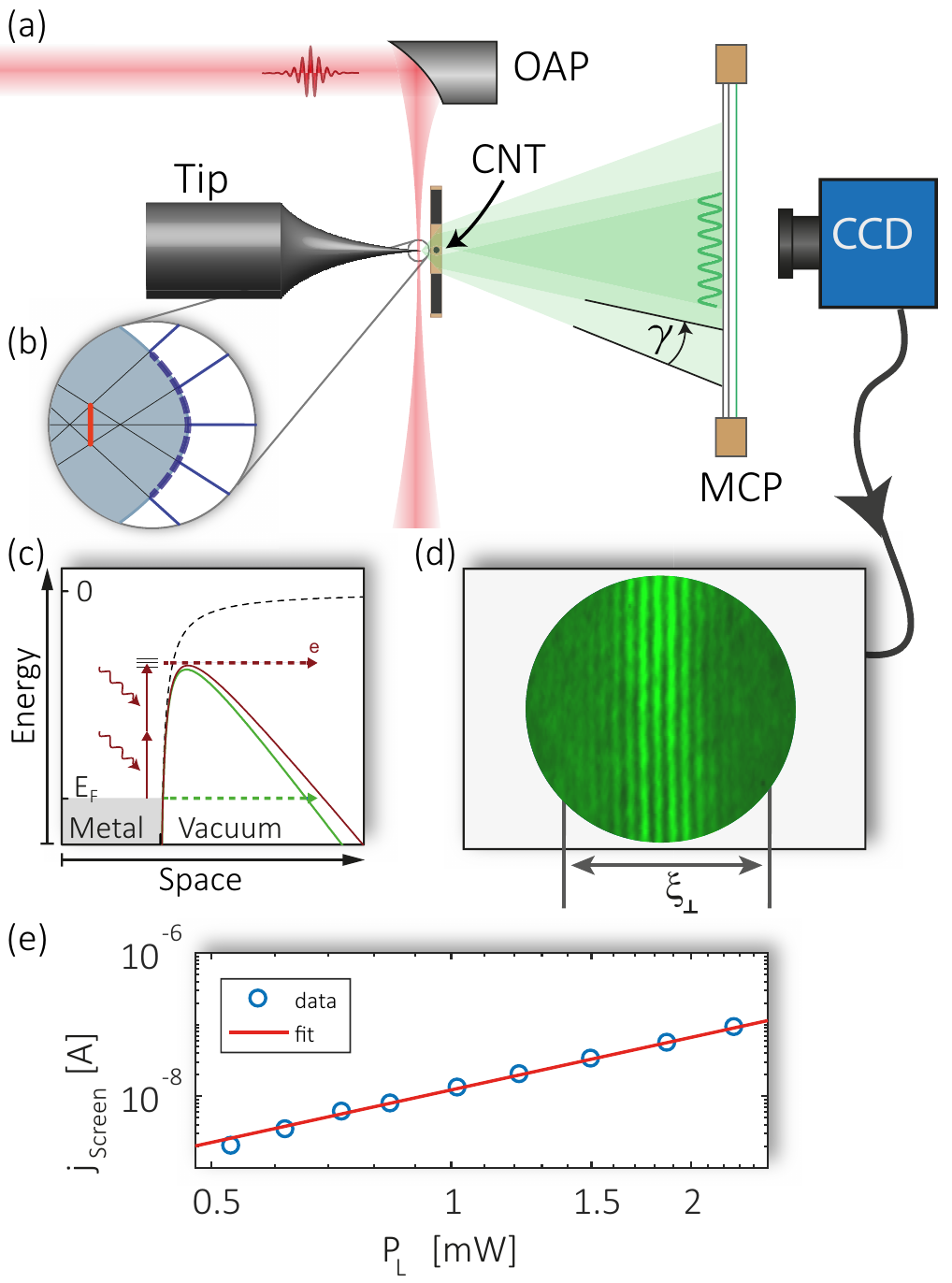}
		\caption{\label{Figure1} 
			(a) Schematics of the experimental setup: near-infrared laser pulses are focused with an off-axis parabolic (OAP) mirror onto a tungsten nanotip, triggering electron emission. A carbon nanotube in front of the tip acts as electron beamsplitter. The emitted electrons are detected by a microchannel plate (MCP) detector, which is recorded with a CCD-camera. (b) Visualization of the effective source size. Backtracking of the emitted electrons inside the tip perpendicular to the tips surface leads to an effectively smaller area than the physical emission area. (c) Different electron emission mechanisms: multiphoton photoemission (red dashed line) and DC-field emission (green dashed line) for a lowered potential barrier. The solid lines show the potential barrier for the respective process. $E_\mathrm{F}$ indicates the Fermi energy, $E=0$ the vacuum. (d) CCD-image of the MCP screen. Electron interference fringes are visible. (e) Measured and MCP-enhanced screen current $j_\mathrm{screen}$ over incident laser power $P_\mathrm{L}$ on the tip. The power law behavior clearly indicates a multiphoton-photoemission process.
		}
	\end{center}
\end{figure}

A quantitative approach to determine the spatial coherence of an electron source is on the basis of spatial interference patterns formed after an electron beamsplitter. From the interference image, one can deduce a figure of merit well suited for the spatial coherence of a source, which is known as the effective source size (Fig.~\ref{Figure1}(b)). The effective source size describes the radius of the smallest area the electrons seemingly originate from when their trajectories are traced back to behind the geometrical emission area.

The van Cittert-Zernike theorem relates the far-field interference fringe pattern with the effective source size \cite{Zernike1938}. It states that the interference width $\xi_\perp$ of a beam, given by the maximum lateral distance at which fringes are still visible, behaves inversely proportional to the effective radius of the emitting area.
The effective source size hence reads \cite{pozzi1987,Spence1993,Cho2004,Ehberger2015}:
\begin{equation}
r_\mathrm{eff}= \frac{\lambda_\mathrm{dB}\cdot l}{\pi\cdot\xi_\perp},
\label{eq1}
\end{equation}
with the distance $l$ between biprism and screen, the de Broglie wavelength $\lambda_\mathrm{dB}=12.3\cdot\sqrt{\left|U_\mathrm{tip}\right|}^{-1}\sqrt{V}\mathring{\mathrm{A}}$ and the lateral width of the far-field interference pattern $\xi_\perp$. $U_\mathrm{tip}$ is the acceleration voltage.

Carbon nanotubes (CNTs) work as excellent beamsplitters in the form of a nano-scale biprism for electrons \cite{Cho2004,Cumings2004,0957-4484-20-11-115401,hasselbach2010,Hwang2013,Ehberger2015}. With a CNT biprism beamsplitter, the effective source size of a tungsten nanotip operated in DC-field emission was measured
to $r_\mathrm{eff}=0.4-0.7\,$nm~\cite{Cho2004}. With blue picosecond pulses (395\,nm) and a blue continuous-wave (cw) beam (405\,nm), a similar experiment has been performed with electrons emitted from a tungsten tip via one-photon photoemission. The resulting interference pattern showed that the photo-emitted electrons form an almost as coherent beam as the DC-field emitted electrons, although the emission processes are physically different. An upper limit for the effective source size of $r_\mathrm{eff}=(0.80\pm0.05)\,$nm was found in this regime \cite{Ehberger2015}.

In this work we present our results on quantitative measurements of the spatial coherence of electrons emitted by few-cycle laser pulses from a Ti:Sapphire oscillator with a central wavelength at 780\,nm. The laser pulses with a duration of 6.1\,fs (FWHM of the intensity envelope) and a repetition rate of $f_\mathrm{rep}=80\,$MHz are obtained from a commercial laser oscillator (Venteon). They are focused on the tungsten tip with an off-axis parabolic mirror with a focal length of $15\,$mm (Fig. \ref{Figure1}). The $1/\mathrm{e}$ spot radius at the tip is $1.1\,\mu$m, resulting in a fluence of $8.1\,\mathrm{mJ}/\mathrm{cm}^2$ at an average laser power of \(12.3\,\)mW (pulse energy of 0.15\,nJ).

The interaction of femtosecond laser pulses and tungsten tips in the given parameter range has been investigated extremely well. For this parameter range, we know from previous work that multiphoton photoemission (MPP) is expected, which is a pertubative and prompt process \cite{Hommelhoff2006,Hommelhoff2006_2, Ropers2007,Yanagisawa2009,Schenk2010,Krger2011,Kealhofer2012,Herink2012, Juffmann2015,Bormann2010,Musumeci2010}. Prompt means that the emission duration of the electron pulse directly reflects the pulse duration of the driving laser pulse and can even be shorter because of the non-linear nature of multiphoton photoemission.

In order to show that MPP takes place, we measure the laser power dependence of the electron beam current $j_n$. An \textit{n}-th order MPP-emission process scales with the laser intensity $I$ as $j_n\sim I^n$ \cite{Boyd2008,Schenk2010,Dombi2010,Krger2018}. Fig.~\ref{Figure1}(e) shows a slope of 2.4 in a double-logarithmic plot, implying that electron emission takes place in an MPP process of order 2 to 3, which is consistent with a mean photon energy of 1.5\,eV and a Schottky-lowered effective barrier height of $3.2$\,eV. Hence, based on the large literature around femtosecond laser triggered tungsten tips \cite{Hommelhoff2006,Hommelhoff2006_2, Ropers2007,Yanagisawa2009,Schenk2010,Krger2011,Kealhofer2012,Herink2012, Juffmann2015,Vogelsang2015}, we conclude that ultrashort electron pulses with a duration reflecting the drive pulse duration are generated.

Tungsten nanotips are electrochemically etched from a tungsten wire grown in the [310]-direction, the direction with the lowest work function of $\phi_\mathrm{[310]}=4.35\,$eV \cite{Mendenhall1937}. A tip is mounted in an ultra-high vacuum chamber with a pressure of about $5\cdot10^{-8}\,$Pa. The geometric tip radius $r_\mathrm{geo}$ was characterized {\it in-situ}  with field ion microscopy~\cite{Mller1960,gault2012atom} to $r_\mathrm{geo}=(6.8\pm1.7)\,$nm.

The nanobiprism beamsplitter consists of a single-walled CNT spanning across a hole on a supporting membrane. The CNTs are grown by chemical vapor deposition (for more information about the growth process, see supplementary information of \cite{hofmann2013})~\cite{Cho2004,hasselbach2010,Ehberger2015}. With a 3-dimensional piezoelectric stage we can approach the CNT to the tip with nanometric resolution. Electrons are detected with a microchannel plate detector (MCP) 6.7\,cm away from the tip.

\begin{figure}[t]
	\begin{center}
		\includegraphics[width=8.5cm]{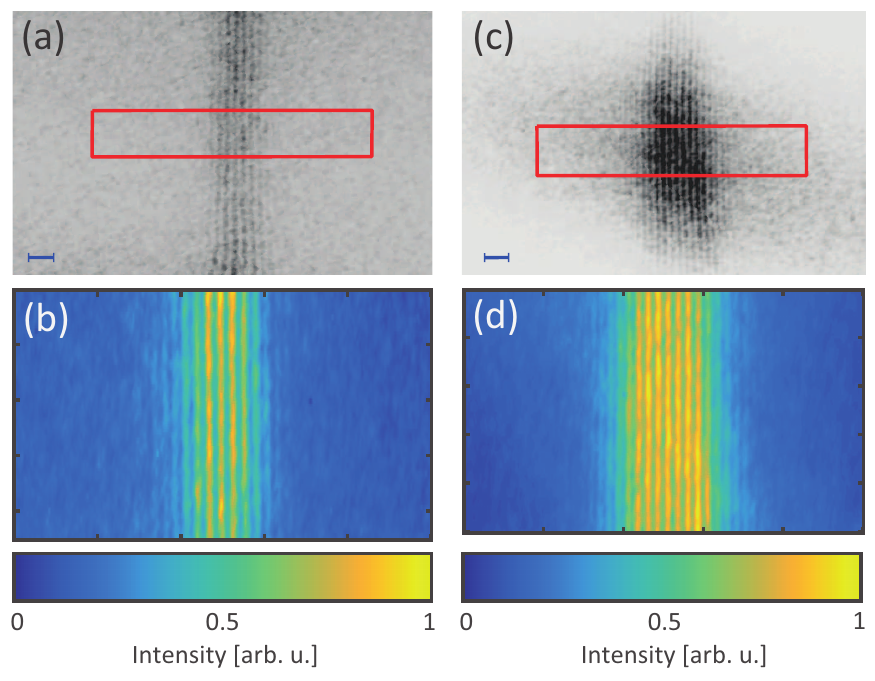}
		\caption{\label{Figure2} 
			(a) CCD-camera image of the MCP detector exhibiting an electron interference pattern. Electrons were emitted in a MPP process. The red frame shows the area of interest. The scale bar is 1\,mm on the screen. (b) Overlap of the red marked region from 30 successively recorded images in MPP. Note that the image is stretched. (c) Interference pattern of electrons emitted by a DC-field emission process recorded under similar experimental conditions like (a). (d) Overlap of the region of interest of 15 images. Colormaps of (a) and (c) are inverted for better visibility.
		}
	\end{center}
\end{figure}

We observe typical interference patterns using multiphoton-photoemitted electrons as shown in Fig.~\ref{Figure2}.
Electron interference fringes are clearly visible. Fig.~\ref{Figure2}(b) shows an integrated image containing 30 frames with an exposure time of 0.663\,s like the one shown in (a) in the red marked area. The positions of the individual images were corrected for drifts and vibrations. The integration of several images enhances the number of visible fringes, as they would blur out in a single image with a longer integration time due to mechanical vibrations.

The bias voltage applied to the tip of $U_\mathrm{tip}=-43\,$V lowers the potential barrier due to the Schottky effect by $\sim1.1\,$eV to $\phi_\mathrm{eff}\approx3.2\,$eV, depending on the surface-dependent field reduction factor \cite{Krger2012}. To prove pure laser-triggered emission, the laser beam was blocked after the measurement resulting in unmeasurable current. For comparison, the bias voltage was set to $U_\mathrm{tip}=-50\,$V in a reference DC-field emission measurement, leading to the comparably large current without laser illumination. Results are shown in Fig.~\ref{Figure2}(c) and (d). In both emission processes one can clearly see electron interference fringes of comparable number and lateral extent.

\begin{figure}[t]
	\begin{center}
		\includegraphics[width=8.5cm]{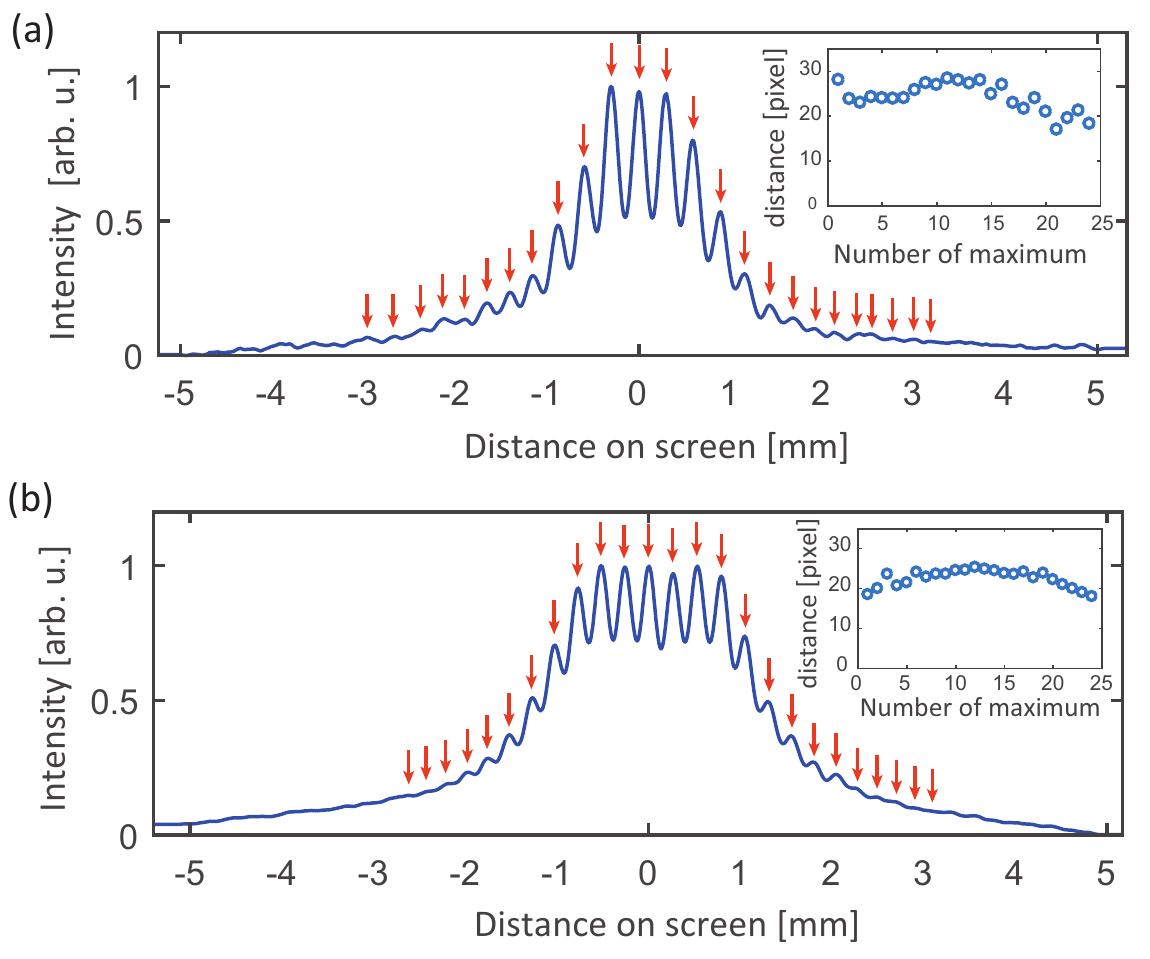}
		\caption{\label{Figure3}
		Evaluation of the data set shown in Fig.~\ref{Figure2}: (a) Recorded line profile for the case of MPP at a bias voltage of $U_\mathrm{tip}=-43\,$V. 25 maxima, marked by the red arrows, were found, with a total interference width of $\xi_\perp^\mathrm{MP}=6.1\,$mm on the MCP screen. (b) Line profile obtained for the case of DC-field emission with a tip voltage of $U_\mathrm{tip}=-50\,$V. 25 Fringes are visible, resulting in an interference width of $\xi_\perp^\mathrm{DC}=5.7\,$mm on the screen (note the slightly lower de Broglie wavelength). The insets in (a) and (b) show the spacing between the maxima. Note that (a) and (b) show curves after subtracting a linear slope of the line profile.
		}
	\end{center}
\end{figure}

To gain a more quantitative insight, we generate line profiles obtained from the datasets of Fig.~\ref{Figure2} by integrating vertically, so perpendicular to the fringe pattern (Fig.~\ref{Figure3}). Except for the bias voltage and laser power, all experimental parameter were kept the same in both measurements for best comparability. In the pattern we obtained for MPP electrons [Fig.~\ref{Figure3}(a)] one can observe up to 25 maxima, resulting in an interference width of $\xi_\perp^\mathrm{MP}=(6.1\pm0.6)\,$mm at a distance of $l=67\,$mm between the CNT and the MCP. With equation~\eqref{eq1} and a de Broglie wavelength of $\lambda_\mathrm{dB}=1.87\,\mathring{\mathrm{A}}$ we derive an upper limit for the effective source size in multiphoton photoemission of $r_\mathrm{eff}^\mathrm{MP}\le(0.65\pm0.06)\,$nm.

For comparison, Fig.~\ref{Figure3}(d) shows the line profile retrieved from the DC-field emission interference pattern, showing also 25 visible fringes and an interference width of $\xi_\perp^\mathrm{DC}=(5.7\pm0.4)\,$mm at a voltage of $-50\,$V and a corresponding de Broglie wavelength of $\lambda_\mathrm{dB}=1.74\,\mathring{\mathrm{A}}$. The resulting effective source size for electrons emitted by DC-field emission is $r_\mathrm{eff}^\mathrm{DC}\le(0.65\pm0.04)\,$nm. The contrast in Fig.~\ref{Figure3}(b) is slightly different than in Fig.~\ref{Figure3}(a) because the smaller fringes in Fig.~\ref{Figure3}(b) blur out more easily, likely due to vibrations and the limited resolution of our imaging system. Note that these limitations hardly impair the analysis of the virtual source size.

This shows remarkably that both emission processes, although physically different, show the same effective source size. Hence the spatial coherence properties of the DC electron beam and that of the ultrafast pulsed beam are very similar. In both cases we see that $r_\mathrm{eff}$ is roughly a factor of 10 smaller than the geometric source size of $r_\mathrm{geo}=6.8\,\mathrm{nm}$ implying that the emitted electrons are (partially) coherent~\cite{Cho2004}. In \cite{Latychevskaia2017a} a detailed simulation suggests that the low-pass-filtered intensity distribution of the emission gives additional insight to the emission nature of the source. However the limited quality of our images prevents the application of this analysis to our data. We note that our and previous discussions around electron trajectories yield estimates of the virtual source size, the exact details of which may also depend on the emission process. An important initial step towards answering questions like these has recently been taken in \cite{Tsujino2018}, addressing the question on the transverse coherence quantum-mechanically. We expect that future combined theoretical-experimental work can bring about new insights into the nature of the transverse properties of ultrafast electron emission.

For distances between the tip and the CNT below a few microns the electron wavefunction is split up at the position of the CNT and shows equidistantly spaced interference fringes on a MCP screen, similar to the original electrostatic biprism experiment \cite{Moellenstedt1956}. The supporting membrane and the CNT are on ground potential, but the close vicinity of the negatively biased tip leads to a bending of the potential around the CNT \cite{Weierstall1999}, leading to the biprism-like behavior. If the CNT is further away from the tip the resulting pattern is dominated by Fresnel diffraction effects of the electrons passing the CNT. In contrast to the equidistant interference fringes in the biprism regime, in this regime the fringe spacing decreases with increasing distance from the center of the pattern. The insets in Fig.~\ref{Figure3} show that the fringe spacing as function of fringe position is almost constant in the center, indicating the biprism action of the CNT. However, also Fresnel diffraction contributions at the outer parts of the pattern show up. For the calculation of the effective source size we take both contributions into account~\cite{Spence1993,Ehberger2015}. We therefore receive a fringe pattern where the distance between the fringes is constant in the center part (biprism effect) and decreases further away from the center (Fresnel diffraction). The outermost fringes we take into account therefore have to follow this slope of the fringe spacing. This way, we make sure we do not identify experimental noise as additional fringes, which would lead to an overestimation of the interference width.

We also consider effects of a finite {\it longitudinal} coherence length as a possible limitation of the measured interference pattern. Finite longitudinal coherence is caused by a finite energy width $\Delta E$ of the emitted electrons. With a known $\Delta E$ we can calculate the expected number of visible fringes. Electrons emitted from room-temperature tungsten needles in DC-field emission display an energy width of $\Delta E=0.3\,\mathrm{eV}$ (full width at half maximum, FWHM) or below~\cite{Ogawa1996,Kiesel2002,Spence2013}. The few-cycle laser pulses used for electron emission have a bandwidth of $\Delta\lambda=430\,\mathrm{nm}$ at -10\,dBc (relative to the peak of the laser spectrum), resulting in an electron energy spread of $\Delta E\approx0.8\,\mathrm{eV}$ (FWHM), consistent with measured MPP-electron spectra~\cite{Schenk2010}. With the energy spread of $\Delta E\approx0.8\,\mathrm{eV}$ we can expect $2n=2E_\mathrm{kin}/\Delta E\approx106$ interference fringes for a kinetic energy of $E_\mathrm{kin}=43\,$eV~\cite{Lichte2008}. We observe a maximum number of 25 fringes in the experiment using MPP-electrons so we conclude that the finite longitudinal coherence is not the limiting factor of the visibility in our measurements, yet it is more limiting than for DC-field emitted electrons.

The interference data was taken with an incident current at the MCP of $4.2\,$pA in MPP emission, which equals on average $0.35\,$ electrons per pulse at a laser repetition rate of $80\,$MHz. Hence we can neglect effects arising from Coulomb repulsion for average currents of around one electron per pulse or above~\cite{Cook2016}. In DC-field emission we measured a current of $2.5\,$pA, so the current in both emission processes was comparable.

In conclusion, we found the electrons emitted in a multiphoton-photoemission process to be as coherent as DC-field emitted electrons. Both result in an effective source size of $(0.65 \pm 0.06)$\,nm. Nonlinearly photoemitted ultrafast electron beams with supreme spatiotemporal properties are ideally suited for various applications in ultrafast electron-based imaging methods.\\
~\newline
The authors acknowledge P. Weber for his technical support. This work has been supported in part by the European Research Council (Consolidator Grant NearFieldAtto and Starting Grant No.~336749) and the Deutsche Forschungsgemeinschaft via the grant SFB 953 and the Nanosystems Initiative Munich (NIM). A.~H. also acknowledges support from the Center for NanoScience (CeNS) and LMUinnovativ.


\bibliographystyle{prsty}

\end{document}